\documentclass[12pt]{article}
\usepackage[T1]{fontenc}
\usepackage{newtxtext,newtxmath}
\usepackage[utf8]{inputenc}
\usepackage[linktocpage=true,plainpages=false]{hyperref}
\usepackage{color}
\usepackage{amsmath,amssymb}
\usepackage{cite}
\usepackage{array}
\usepackage[labelsep=period]{caption}
\usepackage[title]{appendix}

\definecolor{linkcolor}{rgb}{0.6,0,0}
\definecolor{citecolor}{rgb}{0,0.6,0}
\definecolor{urlcolor}{rgb}{0,0,0.9}
\hypersetup{colorlinks, linkcolor={linkcolor},citecolor={citecolor}, urlcolor={urlcolor}}

\newcommand{\ls}{\left(}
\newcommand{\rs}{\right)}
\newcommand{\al}{\alpha}
\newcommand{\be}{\beta}

\newcommand{\dd}{\partial}
\newcommand{\de}{\delta}
\newcommand{\m}{\mu}
\newcommand{\n}{\nu}
\newcommand{\ga}{\gamma}
\newcommand{\la}{\lambda}
\newcommand{\ta}{\tau}
\newcommand{\ka}{\varkappa}

\newcommand{\disn}[2]{$$\displaylines{\refstepcounter{equation}%
		\label{#1}\hskip 1em minus 1em #2\hfilneg}$$}
\newcommand{\nom}{\hfil\hskip 1em minus 1em (\theequation)}

\begin{document}
	\title{Modifications of gravity via differential transformations of field variables}
	\author{A.~A.~Sheykin\thanks{E-mail: a.sheykin@spbu.ru}, \ D.~P.~ Solovyev\thanks{E-mail: dimsol42@gmail.com}, \ V.~V.~Sukhanov\thanks{E-mail: vvsukhanov@mail.ru}, \ S.~A.~Paston\thanks{E-mail: pastonsergey@gmail.com}\\
		{\it Saint Petersburg State University, Saint Petersburg, Russia}}
	
	\date{}
	\maketitle

	\begin{abstract}
		We discuss field theories appearing as a result of applying field transformations with derivatives
		{(differential field transformations, DFT)}
		to a known theory.
		We begin with some simple examples of DFTs to see the basic properties of the procedure. In this process the dynamics of the theory might either change or conserve. After that
		we concentrate on the theories of gravity which
		{appear as a result of various DFT applied to general relativity,}
		namely the mimetic gravity and Regge-Teitelboim embedding theory.
		We review main results related to the extension of dynamics in these theories, as well as the possibility to write down the action of a theory after DFT as the action of the original theory before DFT plus an additional term. Such a term usually contains some constraints with Lagrange multipliers and can be interpreted as an action of additional matter, which might be of use in cosmological applications, e.g. for the explanation of the effects of dark matter.
		
		
	\end{abstract}

\section{Introduction}
The  choice of a proper parameterization is arguably one of the most crucial to solving almost all problems in the theory of gravity. Even in the standard metric formulation of general \mbox{relativity (GR)}, there are dozens of coordinate charts for a given metric \cite{schmutzer}. Picking a suitable coordinate system could significantly simplify the analysis of the metric,  even though physical observables must be coordinate-independent.
In addition to the freedom of   choosing the coordinates, there is also a freedom of choosing the variables which describe a gravitational field. Namely, there are many geometrical quantities which can define the geometry of a manifold.

One alternative to the metric is the connection. A formulation of GR in terms of connection, proposed shortly after the appearance of the original metric one, is called Hilbert--Palatini formulation (although, in Palatini's paper, the connection was \textit{not} treated as independent, and the first who did so was Einstein himself \cite{ferraris_pal}). It allows one to lower the order of derivatives in the gravitational action, which   often turns out very helpful, and addresses many other issues in the theory of gravity,   e.g., its quantization \cite{PhysRevD.91.105006} and unification with other theories (for a historical survey, see \cite{statja47} and the references therein; see also \cite{1712.03061}).

Another early attempt to reformulate GR in terms of alternate variables was made by \mbox{Cartan \cite{1810.03872}} and later by Einstein himself (possibly independently, see \cite{Goldstein2003}). They introduced the notion of the tetrad (or vierbein), which has been extensively used in the GR since then. In particular, it allowed {V.~A.~Fock} to generalize the Dirac equation to the case of curved spacetime \cite{fock_tetr}. Moreover, the tetrad approach provides a very convenient way to study  torsion in the so-called Einstein--Cartan theory (see, e.g. \cite{0711.1535}). Later, Newman and Penrose proposed another kind of tetrad approach to the GR, which was named after them and proved itself very powerful in the analysis of the geometrical structure of GR \cite{1711.11381}.

Probably the most frequently used variables in GR besides the metric are canonical variables of some kind. Indeed,
almost any attempt of quantization requires a Hamiltonian and the corresponding canonical formulation. Since the appearance of the pioneering work of Arnowitt, Deser and \mbox{Misner \cite{adm},} many other canonical formulations of gravity have been developed \cite{0710.4953}. Of course, the canonical approach to GR can be combined with those above   \cite{1604.07764}, leading, e.g., to the loop variables. Besides quantization, Hamiltonian formulation of GR can be of use, e.g., in the analysis of compact \mbox{sources \cite{1805.07240}.}

It must be stressed that all these approaches (at least in their simplest forms) in general do not necessarily lead to the modification of dynamics. However, there are some redefinition procedures of the field variables in gravity, which are by default altering the dynamics of the theory. Many of these are characterized by the presence of additional derivatives of new field variables in the relation between old and new ones.

The main purpose of this paper is to put together the results and methods related to the extended dynamics of Lagrangian theories with differential field transformations (DFT). The paper is organized as follows. In    Section \ref{secII}, we
discuss
the basic properties of DFTs in several toy examples. \mbox{In    Section \ref{secIII}}, we focus on the first historical example of DFT in gravity, namely the Regge--Teitelboim gravity, \mbox{in which} the extension of dynamics occurs, so it can be of use in cosmological applications.
In    Section \ref{secIV}, we consider   another cosmologically relevant example of DFT that has drawn some attention in   recent years: the mimetic gravity. We also briefly discuss the generalization of the mimetic transformation called disformal transformation.

\section{Examples of Theories with Differential Field Transformations}\label{secII}
\vspace{-6pt}

\subsection{Mechanics}\label{2.1}
Let us consider a toy model \cite{statja33} that represents some of the essential features of theories with differentially transformed fields. Namely, having the action of a harmonic oscillator:
\begin{align}
S=\int dt \left(\frac{\dot{q}^2}{2}-\frac{\omega^2{q}^2}{2}\right) \label{osc1}
\end{align}
the corresponding equations of motion (EoM) has the form
\begin{align}
\ddot{q}+\omega^2 q = 0. \label{osc1a}
\end{align}

If one performs a substitution
\begin{align}\label{y}
q(t)=\dot{y}(t)
\end{align}
then
the action in   Equation      \eqref{osc1} takes the form
\begin{align}
S=\int dt \left(\frac{\ddot{y}^2}{2}-\frac{\omega^2{\dot{y}}^2}{2}\right) \label{osc1b}
\end{align}
and as a result of applying of the Leibnitz rule to the variation
\begin{align}\label{osc2b}
\delta S=-\int dt \left(\dddot{y}+\omega^2\dot{y}\right)\delta \frac{d}{dt} y  = \int dt \frac{d}{dt}\left(\dddot{y}+\omega^2\dot{y}\right)\delta  y = 0
\end{align}
the following EoM appears:
\begin{align}
\frac{d}{dt}\left(\dddot{y}+\omega^2\dot{y}\right) = 0 \label{osc3}.
\end{align}
{This can be integrated to obtain}
\begin{align}
\dddot{y}+\omega^2\dot{y} = C, \label{osc3a}
\end{align}
and transformed back to the old variables:
\begin{align}
\ddot{q}+\omega^2 q = C, \label{osc3b}
\end{align}
which is not equivalent to Equation      \eqref{osc1a}.

The equivalence is lost because the substitution in  Equation      \eqref{y} changed the class of variations in the variational principle  in  Equation      \eqref{osc2b}.
Indeed, fixing the endpoints in the original variational principle implies that
\begin{align}\label{dq}
{\delta q(t_{1,2}) = 0,}
\end{align}
whereas in the Lagrangian      (Equation      \eqref{osc1b}) second derivatives are present, and one must require
that \mbox{not only}
{ \begin{align}\label{dy}
	\delta y(t_{1,2}) = 0,
	\end{align}
	but also
	\begin{align}\label{ddoty}
	\delta \dot{y}(t_{1,2}) = 0
	\end{align}  }
holds.

The condition  in  Equation      \eqref{ddoty} is equivalent to the usual fixation of endpoints in Equation      \eqref{dq} in the original variational principle, but if we interpret Equation      \eqref{dy} in terms of this principle, it turns out that we use a restricted class of variations $\de q$, which satisfies  the condition
\begin{align}\label{dy2}
\int_{t_1}^{t_2} dt\,\delta q = 0.
\end{align}

Thus,    the substitution in  Equation      \eqref{y} leads to the contraction of the class of {variations}
in the variational principle,
which gives us  fewer  restrictions on the dynamics of $y(t)$ itself
\cite{Golovnev201439}.

However, the extension of dynamics after   differential transformations of the original fields is not guaranteed. The simplest mechanical illustration of the conservation of dynamics can be found in \cite{1702.01849}. The authors considered  the Lagrangian
\begin{align}\label{meh2-1}
L = \frac{\dot{X}^2}{2}+\frac{\dot{Y}^2}{2}
\end{align}
and the substitution
\begin{align}\label{meh2-2}
X = x-\dot{y}, \ Y=y.
\end{align}

Since this substitution is invertible and can be uniquely solved with respect     to  new variables $x$ and $y$, the resulting Euler--Lagrange equations on $x$ and $y$ turn out to be of second order,  thus they require the same amount of initial values as original ones, and the dynamics remains unchanged.

On the contrary, the set of solutions of Equation      \eqref{osc3b} is larger than that of Equation      \eqref{osc1a}: if $C\neq 0$, some {``extra solutions'' }appear due to the lack of invertibility of the substitution in  Equation      \eqref{y}. These extra solutions of Equation      \eqref{osc3b} are governed by the value of a single constant, which is, in turn, defined by the initial values of the new variable $y$,  thus, if at any moment $t_0$ these values  are chosen in the way that $C=0$, then $q(t)$ will always solve Equation      \eqref{osc1a}.

It must be stressed that here and hereafter we   use the concept of conservation of dynamics in the following sense: \textit{The dynamics of the theory after the field transformation is conserved if the amount of initial values necessary to solve the EoM of the theory in the new variables coincide with the amount of initial values in the original ones.} If this amount is larger than the original one, we   refer to this case as the case of extended dynamics, and if this amount is smaller, we refer to it as the restricted dynamics.

\subsection{Massive Scalar Field}\label{2.2}
In a field theory, the situation becomes slightly more complicated. Let us consider an action of a massive scalar field
\begin{align}
S=\frac{1}{2}\int d^4 x \left(\partial_\mu \phi\, \partial^\mu \phi -m^2{\phi}^2\right) \label{f1}
\end{align}
and a quadratic differential substitution
\begin{align}
\phi=\partial_\nu \psi\, \partial^\nu \psi  \label{f2}.
\end{align}

Note that linear substitution without any additional fields is no longer possible   since we need to construct the scalar $\phi$ from the vector $\partial_\mu \psi$.  This substitution transforms the original EoM
\begin{align}
\square\phi+m^2\phi=0  \label{f3}
\end{align}
to
\begin{align}
\partial_\mu\ls(\square\phi+m^2\phi)\partial^\mu \psi\rs=0  \label{f4}.
\end{align}

This equation can be interpreted as a conservation law
\begin{align}
\partial_\mu  j^\mu = 0  \label{f5}
\end{align}
of some current
\begin{align}
j^\mu = (\square\phi+m^2\phi) \partial^\mu \psi. \label{f5a}
\end{align}

To rewrite Equation      \eqref{f4} in a form similar to Equation      \eqref{f3}  (by analogy with the above example, where Equation      \eqref{osc3} is rewritten in the  form  of Equation      \eqref{osc3b} close to Equation      \eqref{osc1a}), let us denote
$J\equiv  \square\phi+m^2\phi$. Then, it is possible to rewrite Equation      \eqref{f4} as the following system:
\begin{align}\label{f8a}
\square\phi+m^2\phi=J,\\
\partial_\mu (J \partial^\mu \psi) = 0,\label{f8b}
\end{align}	
thus we have $j^\mu = J \partial^\mu \psi$ as an expression for the conserved current. It can be seen that,
as a result of the substitution in  Equation      \eqref{f2}, the extension of dynamics occurs. Note that,
for field theory, in contrast with the mechanical system mentioned above, extra solutions are governed not by a single constant, but rather by the new
dynamical variable $J(x)$, and the solutions of the original theory correspond to $J(x)=0$.

It is interesting that    Equations      \eqref{f8a} and       \eqref{f8b}, as well as the substitution in  Equation      \eqref{f2}, can be obtained simultaneously as EoM of the following action:
\begin{align}
S=\int d^4 x \left(\frac{1}{2}\ls\partial_\mu \phi\, \partial^\mu \phi-m^2{\phi}^2\rs + J (\phi-\partial_\mu \psi\, \partial^\mu \psi)\right), \label{f9}
\end{align}
in which $\phi$, $\psi$ and $J$ are treated as independent variables. This action is obtained by adding the constraint to the original action in   Equation      \eqref{f1},
in which $J$ plays the role of a Lagrange multiplier (\mbox{the variation} with respect     to  it gives Equation      \eqref{f2}). Therefore, the theory can be considered as a theory of the original field $\phi$ interacting with additional fields $\psi$ and $J$.
Note that such alternative description of the theory appearing as a result of DFTs can be used in more complicated cases as well, e.g., in gravity  (see Section \ref{secIII}).

\subsection{The Hilbert--Palatini Approach}\label{2.3}
The Hilbert--Palatini approach, which is very familiar in the GR, can also serve as an example of theory with differential relation between fields. To make it similar to the above examples, let us start with the action
as a sum of the Einstein--Hilbert (EH) action with independent connection and a material contribution $S_m$:
\begin{align}\label{hp}
S = -\frac{1}{2\varkappa}\int d^4 x \sqrt{-g}\, g^{\mu\nu}R_{\mu\nu} (\Gamma) +S_m.
\end{align}

In this action, we have 10 metric functions $g_{\mu\nu}$ and 40 components of \textit{symmetric} connection $\Gamma^\al_{\mu\nu}$, which gives 50 independent variables in total. The EoM of this action are the usual Einstein equations
\begin{align}
\frac{\delta S}{\delta g_{\mu\nu}} = 0 \quad\Rightarrow\quad G^{\mu\nu}  - \varkappa\, T^{\mu\nu}=0
\end{align}
and { the equation appearing as a result of variation with respect     to  $\Gamma$}
\begin{align}
\frac{\delta S}{\delta \Gamma^\al_{\mu\nu}} = 0 \quad\Rightarrow\quad \label{chris}
\Gamma^{\alpha}_{\mu\nu}  =\frac{1}{2} g^{\alpha\beta}(\partial_\mu g_{\nu\beta}+\partial_\nu g_{\mu\beta} - \partial_\beta g_{\mu\nu}),
\end{align}
which is the usual condition of consistency of the connection with the metric corresponding to Riemannian geometry. 

Now,  let us consider the consistency condition, i.e.,   Equation      \eqref{chris} as DFT, as a result of which \mbox{50 independent} variables of the theory in Equation      \eqref{hp} become expressed through 10 variables of a new theory, namely the metric components $g_{\m\n}$. From the above examples, one can expect that substitution of this DFT
back into the action will alter the dynamics.
However, it is obvious that the resulting theory is the original GR.
Therefore, in this case, the extension of dynamics does not occur. The reason  for  this lies in the fact that the chosen DFT coincides with one of the EoM of the \mbox{original theory.}

In this example, one can write down the action of the new theory by analogy with the consideration in the previous subsection, adding a DFT with the Lagrange multiplier $J_{\al}{}^{\mu\nu}$ (symmetric by $\mu\nu$)  to the original action in   Equation      \eqref{hp}:
\begin{align}\label{hpn}
S = -\frac{1}{2\varkappa}\int d^4 x \sqrt{-g}\ls g^{\mu\nu}R_{\mu\nu} (\Gamma)+
J_{\al}{}^{\mu\nu}\ls \Gamma^{\alpha}_{\mu\nu}  -\frac{1}{2} g^{\alpha\beta}(\partial_\mu g_{\nu\beta}+\partial_\nu g_{\mu\beta} - \partial_\beta g_{\mu\nu})\rs
\rs+S_m.
\end{align}

In this theory, the original fields $g$ and $\Gamma$ interact with additional field $J$.
It is easy to check that the EoM corresponding to this action lead to the condition $J_{\al}{}^{\mu\nu}=0$ due to its specific structure, which means the conservation of dynamics after DFT.

The conservation of dynamics in the considered
case is related to the commutativity of this particular substitution in  Equation      \eqref{chris} in the action and in the EoM \cite{0909.4151}. It is worth noting that, if the connection in the original action in   Equation      \eqref{hp} is not  {a priori} symmetric, the situation drastically changes:
the original theory is not equivalent to GR anymore and contains additional degrees of freedom \cite{statja47}, whereas DFT transforms the theory into GR, i.e., we arrive at a case of the \mbox{restricted dynamics.}

From this example, we can see that it is not sufficient to apply a differential transformation of fields to the EH action to obtain an extended theory. Nevertheless, this kind of tricks proves itself useful in the theory of gravity, and in the next section we   discuss the concrete examples of extended dynamics after {DFT}.

\section{Isometric {Embeddings} and Regge--Teitelboim {Gravity}}\label{secIII}
One of the first historical appearances of the geometric variables, which are connected to the metric through differential relations, occurred in a procedure of isometric embedding. In this procedure a pseudo-Riemannian spacetime is considered as a surface in an ambient spacetime of higher dimension, which can be pseudo-Riemannian as well, but is often assumed to be flat (or Ricci-flat \cite{gr-qc/0111058}, \mbox{or conformally} flat \cite{Dunajski_2019} ). The number of ambient spacetime dimensions required for \emph{local} isometric embedding is governed by Janet--Cartan--Friedman theorem \cite{gane,kart,fridman61}. For a generic $n$-dimensional pseudo-Riemannian spacetime, the number $N$ of ambient spacetime dimensions is
\begin{align}
N\geq \frac{n(n+1)}{2}.
\end{align}

If a global embedding is considered, this number dramatically increases. For example, $N\geq 48$ in the case of generic compact four-dimensional spacetime embedded in a flat ambient spacetime and $N\geq 89$ in the case of a non-compact one \cite{clarke}. However, most symmetric spacetimes can be embedded (even globally) in the spacetimes with a relatively small  $N$; e.g., for a {Friedmann--Robertson--Walker (FRW)} model $N=5$ {(see, e.g., \cite{statja29})} and for a spherically symmetric spacetime $N=6$ \cite{statja27}.

The main variable which defines the external geometry of the surface is the embedding function $y^a(x^\mu)$. It is connected to the metric through the inducibility condition
\begin{align}
g_{\mu\nu}(x) = \partial_\mu y^a(x) \partial_\nu y^b(x) \eta_{ab},\label{met}
\end{align}
which follows from the fact that the embedding is isometric and
\begin{align}
ds^2=\eta_{ab} dy^a(x) dy^b(x) = g_{\mu\nu} dx^\mu dx^\nu.\label{met1}
\end{align}

Here, $\eta_{ab}$ is the metric of the ambient space, which here and hereafter is assumed to be flat. Note that there is no restriction on the number of   spacelike and timelike dimensions of the ambient spacetime, provided they are equal to or greater than the numbers of the timelike and spacelike dimensions of the embedded spacetime,  thus an embedding of $(3+1)$-dimensional spacetime in $(5+1)D$, $(4+2)D$ or $(3+3)D$ spacetime is equally possible.

The area of applications of isometric embeddings is quite large and includes the study of the geometric structure of the Riemannian manifolds \cite{goenner}, the classification of solutions of the Einstein equations \cite{schmutzer}, the calculation of the thermodynamic properties of the spacetimes with a horizon \cite{deserlev99,statja36}, the definition of energy in various theories of gravity \cite{statja46,0805.1370}, and so on.

Among them, there are some attempts to construct a modified theory of gravity on the basis of EH action in which the substitution in  Equation      \eqref{met} has made and the embedding function is treated as an independent dynamical variable instead of metric.
Such a procedure exactly corresponds to the procedure of DFTs, which is discussed here.
Regge and Teitelboim were the first to propose  such an idea in 1975 \cite{regge},  thus the corresponding theory is often called the Regge--Teitelboim
{gravity} (\mbox{or the} embedding theory), although in the subsequent years it had been independently rediscovered many times  (see \cite{tapiaob} for a list of references and also \cite{estabrook2009}). The initial interest to this theory was quite small since in the original paper it was incorrectly stated that the constraint algebra of the theory is not closed due to a calculation error which was corrected only in 2007 \cite{statja18}.
Moreover, in the 1970s and 1980s, the appearance of the extra solutions has to be treated as a drawback of the theory since there was no big  evidence that the Einsteinian dynamics cannot explain any of the observational \mbox{phenomena \cite{deser}.}
However, since the 1990s, an interest in this approach has been growing, as  it turns out that extra solutions can simulate the effects of dark matter and/or dark energy.


The basis of the Regge--Teitelboim approach is the Einstein--Hilbert action with a matter:
\begin{align}\label{EHM}
S = -\frac{1}{2\varkappa}\int d^4 x \sqrt{-g}\, g^{\mu\nu}R_{\mu\nu} +S_m
\end{align}
(although it might be generalized on the case of an arbitrary metric theory, see. e.g., \cite{kokarev1998}).
If the DFT   in Equation      \eqref{met} is made (to keep the number of fields unchanged, one needs to put $N=10$, although other choices were also considered in the literature \cite{bustamante}), the variational principle gives
\begin{multline}
\delta S = \frac{1}{2\varkappa} \int d^4 x  \sqrt{-g}\, (G^{\mu\nu} - \varkappa T^{\mu\nu}) \delta g_{\mu\nu}=\\
= \frac{1}{\varkappa} \int d^4 x  \sqrt{-g}\, (G^{\mu\nu} - \varkappa T^{\mu\nu}) \eta_{ab} \partial_\mu y^a \partial_\nu \delta y^b  = \\ =
-\frac{1}{\varkappa} \int d^4 x\,  \partial_{\mu}\ls \sqrt{-g}\, (G^{\mu\nu} -\varkappa T^{\mu\nu} ) \partial_{\nu} y^a\rs \delta y_a=0,	\label{var_RT}
\end{multline}
thus the EoM have the following form:
\begin{align}
\partial_{\mu}\ls \sqrt{-g} (G^{\mu\nu} -\varkappa T^{\mu\nu} ) \partial_{\nu} y^a\rs =	0,\label{RT}
\end{align}

They are called Regge--Teitelboim equations (RT). Note the complete analogy of the structure of   Equations      \eqref{RT} and        \eqref{f4}. This analogy can be extended further if one notes that RT equations can be rewritten as a conservation law of some set of currents:
\begin{gather}
\partial_{\mu} j^\mu_a =	0,\label{div}\\ j^\mu_a= \sqrt{-g} (G^{\mu\nu} -\varkappa T^{\mu\nu} ) \partial_{\nu} y_a \label{curr}.
\end{gather}

Moreover, they can be transformed into the Einstein equations with the additional contribution $\tau^{\mu\nu}$ in the energy-momentum tensor (EMT):
\begin{align}\label{RT_Ein}
G^{\mu\nu} =\varkappa (T^{\mu\nu}+\tau^{\mu\nu}),
\end{align}
where
\begin{align}\label{tau}
\tau_{\mu\nu} = \frac{j_\mu^a \partial_{\nu} y_a}{\varkappa\sqrt{-g}}
\end{align}
is a new dynamical variable whose dynamics is governed by the equation
\begin{align}
\partial_{\mu}( \sqrt{-g} \tau^{\mu\nu}\partial_{\nu} y^a) = 0,\label{RT_}
\end{align}
which is analogous to     Equation      \eqref{f8b} and has a similar structure.

We see that, in this case, which is analogous to the one considered in Section  \ref{2.2}, the extension of dynamics occurs as a result of the DFT   in Equation      \eqref{met}.
Furthermore, as in Section      \ref{2.2}, the dynamics coincides with the original one if $\tau^{\mu\nu}=0$ in an arbitrary moment of time (or at arbitrary value of any coordinate, see \cite{statja33}). However, it was shown in  \cite{statja18} that the conservation of dynamics can be achieved {(with certain technical assumptions)} by imposing a much weaker condition
\begin{align}\label{eincon}
\tau^{0\mu} = \frac{1}{\ka}\ls G^{0\mu}-\varkappa T^{0\mu}\rs=0
\end{align}
only at the initial moment of time.
This possibility of eliminating of the extra solutions through imposing of the restriction at the initial moment of time is analogous to the possibility to do it for the mechanical example in Section \ref{2.1}, when one sets the initial values in such a way that $C=0$ (see the discussion at the end of Section \ref{2.1}).

The set of conditions in Equation      \eqref{eincon} is called Einsteinian constraints, and in the original formulation of Regge--Teitelboim approach these were imposed to ensure that the dynamics of gravity in terms of embedding variables remains the same as in the standard metric approach. The reason for this lies in the motivation of Regge and Teitelboim: in the development of this approach, they did not want to modify gravity, but rather to find some alternative variables which would be more suitable for the quantization purposes.
The canonical formulation of this approach with an additional imposing of Einsteinian constraints, including the explicit form of the full constraint algebra, was obtained in  \cite{statja24}.
Without the Einsteinian constraints, the dynamics of the theory becomes less restricted, the canonical formulation becomes drastically more complicated since the appearing constraints can be written only in implicit form. However, even in this case, the canonical formulation together with the constraint algebra can be obtained \cite{statja44}.

In later years, the motivation for studying of embedding theory has shifted towards the cases of extended dynamics, namely the ones which allow the cosmological applications.
{There was an attempt to explain effects that were usually treated in GR as effects of dark energy presence, as extra solutions of the embedding theory \cite{davids97}.
	The attempts to explain the effects related to the problem of dark matter in GR should be considered as more successful. First such consideration at the cosmological scales was done by   \cite{davids01}. However, it turned out that such an approach	
	requires fine-tuning of the initial conditions of the universe \cite{statja26}.
	There might be a possibility of using extra solutions in the description of the inflationary phase of the universe, although we are not aware of any successes in this field.
	
	In the analysis of the extra solutions, an important role is played by a choice of background embedding function.
	In cosmological applications, authors usually consider the simplest five-dimensional embedding of a FRW model, which  is mentioned above (it is this embedding that was used in \cite{davids97,davids01,statja26}). Due to a small number of ambient space dimensions in which the chosen background embedding function is embedded,
	in the perturbative expansion, the RT equations turn out to be nonlinear with respect     to  perturbations. Therefore, the results obtained in the cosmological approximation (i.e.,   in the assumption of the homogeneity and isotropy of space) may not be relevant. A better (and possibly physically based) choice of a background embedding function can help one to choose a suitable set of independent variables in the description of embedding theory. At the same time, a reformulation of the theory in new variables might ease the analysis of the EoM of the theory. The role of such variables can be played by the quantities $\ta^{\m\n}$ (in this case the EoM are Equations      \eqref{RT_Ein} and      \eqref{RT_}) or $j^\m_a$.
	
	Recently, such reformulations of embedding theory were developed, in which the geometric interpretation (possessed by embedding theory in the original variables $y^a$ {only})  is sacrificed in favor of the physical one.
	The first of them \cite{statja48} appears as a result of the exactly same procedure that transformed Equation      \eqref{f1} into Equation      \eqref{f9} {(as well as Equation      \eqref{hp} into Equation      \eqref{hpn})}:
	\disn{v2a}{
		S[\phi(\psi)] \to S[\phi]+S^{\text{add}}[\phi,\psi,J],
		\nom}
	where $S$ is the original action of the theory, $\phi$ are the old variables, $\psi$ are new ones and $J$ are Lagrange multipliers which correspond to the relations between $\phi$ and $\psi$. Namely, let us add an expression with DFT, namely the inducedness condition
	\disn{za1}{
		S^{\text{add}}=\frac{1}{2}\int\! d^4 x\, \sqrt{-g}\,
		\Bigl( (\dd_\m y^a)(\dd_\n y_a) - g_{\m\n}\Bigr)\tau^{\m\n},
		\nom}
	with the Lagrange multiplier $\tau^{\mu\nu}$ to the original action in   Equation      \eqref{EHM}.
	It is easy to check that it gives the RT equations in the form of Equations      \eqref{RT_Ein} and      \eqref{RT_}.
	
	If one wants to use the set of currents $j^\m_a$ as independent variables, one should use the following form of the additional term in action \cite{statja51}:
	\disn{r5}{
		S^{\text{add}}=\int\! d^4 x\, 
		\Bigl( j^\m_a\dd_\m y^a-\text{\bf tr}\sqrt{g_{\m\n}j^\n_a j^{\al a}}\Bigr),
		\nom}
	which cannot be reduced to the used DFT with the Lagrange multiplier. Note that a Lagrange multiplier is nevertheless contained in   Equation      \eqref{r5}. It is easy to see that its role is now played by the embedding function $y^a$. Variation with respect     to  $y^a$ gives the conservation law for the currents in  Equation      \eqref{div}. In Equation      \eqref{r5},
	\text{\bf tr} means trace of a square root of matrix with indices $\m$ and $\al$ 
	(see the discussion of the details of the matrix root usage in the description of gravity in \cite{golovnev2017}).
	It can be checked (see details \mbox{in \cite{statja51}}; note that $j^\m_a$ used in the present paper differs from the analogous quantity in \cite{statja51} by a multiplier $\sqrt{-g}$) that EoM of such theory are Equations      \eqref{div} and      \eqref{curr}
	and the inducedness condition  in  Equation      \eqref{met}.
	
	The usage of (on-shell) conserved currents $j^\m_a$ seems the most promising, since one can interpret them as a characteristics of the flow of some additional \textit{embedding matter} \cite{statja51}.
	In the analysis of its properties, one can use the similarity between its action in
	Equation      \eqref{r5}
	and the action of the additional \textit{mimetic matter}, which appears in the investigation of the another gravitational theory with DFT, namely the mimetic gravity.}

\section{Mimetic Gravity and Its Extensions}\label{secIV}
\vspace{-6pt}

\subsection{Mimetic Gravity}\label{4.1}
In contrast with thte Regge--Teitelboim approach, which initially  did not get its deserved credit and nowadays is seen by some as {``old-fashioned'',} the mimetic model for gravity was proposed \cite{mukhanov} several years ago and immediately drew much attention. This approach was motivated by the problem of dark matter.

The main idea of the mimetic gravity is the following. Let us consider the ordinary EH action {with matter in Equation      \eqref{EHM}, and the following DFT:}
\disn{v1}{
	g_{\m\n}=\tilde g_{\m\n}\tilde g^{\ga\de}\dd_\ga\la\, \dd_\de\la,
	\nom}
where $g_{\m\n}$ is a physical metric, $\tilde g_{\m\n}$ is an auxiliary metric and $\la$ is an additional scalar field.
Note that Weyl transformations of the auxiliary metric do not change $g_{\m\n}$,
{thus  the conformal mode of $\tilde g_{\m\n}$ does not affect the action. As a result of DFT, one obtains that
	the ten original variables (the components of the physical metric $g_{\m\n}$) are expressed through nine remaining degrees of freedom of the auxiliary metric $\tilde g_{\m\n}$
	and the scalar field $\la$,  thus the number of DoFs after this DFT is the same, if the conformal mode of $\tilde g_{\m\n}$, which is pure gauge, is not counted.
	This situation is often referred to as the isolation of conformal mode.}

The corresponding system of EoM {has the form
	\begin{gather}
	G^{\m\n} =\ka \ls T^{\m\n}+ n\, g^{\mu\al} g^{\nu\be} \partial_\al \la\, \partial_\be \la\rs,\label{ur1}\\
	\partial_\mu \ls\sqrt{-g}\, n\, g^{\mu\nu}\partial_\nu \la \rs = 0,\label{ur2}
	\end{gather}
	where
	\disn{vi17a}{
		n\equiv g_{\m\n}\ls \frac{1}{\ka}G^{\m\n}-T^{\m\n}\rs.
		\nom}
	
	As can be seen, Equation      \eqref{ur1} is the Einstein equations, i.e., the EoM of the original theory, but with some additional \textit{mimetic} matter.
	The second EoM in Equation      \eqref{ur2} can be interpreted as a conservation of some current
	\begin{gather}
	\partial_{\mu} j^\mu =	0,\label{div2}\\
	j^\mu= \sqrt{-g}\, n\, g^{\mu\nu}\partial_\nu \la  \label{curr2},
	\end{gather}
	i.e., the structure of the EoM in  Equations      \eqref{ur1} and      \eqref{ur2}
	reproduces the structure of EoM in  \mbox{Equations      \eqref{f8a}    and        \eqref{f8b}} from   Section \ref{2.2} and the EoM in Equations      \eqref{RT_Ein}    and        \eqref{RT_} from    Section \ref{secIII}.
}

We see that, as in  Sections \ref{2.2} and \ref{secIII}, as  a result of the DFT   in Equation      \eqref{v1}, the extension of dynamics occurs, and the extra solutions are governed by the variables $n$ and $\la$,  thus
the solutions of the original theory (i.e.,   GR) correspond to $n=0$.
Note that, as can be seen from the structure of  the right-hand side  of Equation      \eqref{ur1}, the quantity $n$ plays the role of concentration of mimetic \mbox{matter particles,}
\disn{vi17b}{
	u_\m=\dd_\m\la
	\nom}
is a 4-velocity of them, and the motion of mimetic particles turns out to be potential one. It must be stressed that, as one can easily check, the condition
\disn{vi18}{
	g^{\m\n}u_\m u_\n=g^{\m\n}\dd_\m\la\, \dd_\n\la=1
	\nom}
is a consequence of the DFT   in Equation      \eqref{v1}.

As in the above examples, for the theory appearing as a result of  the DFT   in Equation      \eqref{v1} in the action in   Equation      \eqref{EHM},
a new form of action can be written analogously to Equation      \eqref{v2a}, i.e., as a sum of Equation      \eqref{EHM} and
additional term $S^{\text{add}}$ depending on new fields, where the physical metric $g_{\m\n}$ is treated as independent variable.
In the variables $\la$ and $n$, it was done {in \cite{Golovnev201439}:}
\disn{v3}{
	S^{\text{add}}=
	\int\! d^4x\sqrt{-g}\, \Bigl(g^{\m\n}(\dd_\m\la)(\dd_\n\la) - 1\Bigr)n,
	\nom}
{where} $n$ plays the role of a Lagrange multiplier.
{Since Equation      \eqref{v3} describes a pressureless perfect fluid with potential motion,
	it is possible to rewrite it in many other forms \cite{statja48}, which can be of use in certain situations. 
	
	It is also worth noting that mimetic DFT can be applied not only to Einstein--Hilbert, but to any metric theory as well, e.g.,   Gauss--Bonnet gravity. The action of resulting theory might be rewritten  using Lagrange multipliers despite   Equation      \eqref{v3}  \cite{1504.04861}. Moreover, there are other modifications of gravitational variables that do not include DFTs but, as   in the case of DFTs,  might be interpreted as GR with additional action terms containing scalar fields. The theories with non-metric measure \cite{1205.1056}, which provides  a unified description of dark energy and dark matter, can serve as an example of such~modification.}

{ To compare the results with embedding theory discussed above, let us write the equivalent form of the action in   Equation      \eqref{v3}, in which the role of independent variable is played by a current $j^\m$ corresponding to  Equation      \eqref{curr2}:}
\disn{p2}{
	S^{\text{add}}=\int\! d^4 x \ls j^\m \dd_\m\la - \sqrt{g_{\m\n} j^\m j^\n }\rs,
	\nom}
where the Lagrange multiplier for the conservation in Equation      \eqref{div2} of the current $j^\mu$
is now given by a scalar $\la$.
In such choice of independent variables, the density of mimetic matter $n$ and its 4-velocity $u^\m$ are expressed through independent variables by formulas
\disn{p3}{
	n= \frac{\sqrt{j^\m j^\n g_{\m\n}}}{\sqrt{-g}},\qquad
	u^\m= \frac{j^\m}{\sqrt{j^\m j^\n g_{\m\n}}}
	\nom}
(see details in \cite{statja48}; note that $j^\m$ used here differs from analogous quantity in \cite{statja48} by the multiplier~$\sqrt{-g}$).

The action in   Equation      \eqref{p2} looks very similar to Equation      \eqref{r5}, { thus  it makes sense to search for the limiting case of the embedding theory in which it reduces to mimetic gravity. It might help in the studying of the properties of embedding matter as the properties of
	mimetic matter is easy to understand: it is a pressureless dust matter with potential motion.
	The idea of such limit was proposed in \cite{statja48}, but the study of this approach has not finished yet.
	
	Note that in the case of embedding theory the motion of {the} embedding matter is already unrestricted,
	thus it is not necessary to modify the theory further.
	On the contrary, the mimetic gravity in its original formulation (Equation      \eqref{vi18}) is too simple and restricted, and, in fact, can only serve as a starting point in various generalizations and modifications,
	{after that in the framework of mimetic gravity arises the possibility to explain certain problems of modern cosmology \cite{mimetic-review17}.
		Among these modifications, a possibility to add the potential for scalar $\la$ was considered \cite{mukhanov2014} (which lead to the appearance of pressure of mimetic matter), higher derivative contributions of this field \cite{mukhanov2014,vikman2015,kobayashi2017}
		(which transforms mimetic matter into imperfect fluid), the introduction of
		{additional scalar fields \cite{Firouzjahi_2018}, non-minimal coupling of mimetic scalar with matter \cite{Vagnozzi_2017}, unification with unimodular gravity \cite{1601.07057}, etc. A comprehensive review of these and other modifications of mimetic gravity connected to the choice of the Lagrangian can be found in \cite{1705.11098}.
			One can also generalize the theory by modifying the expression for the mimetic DFT rather than the  action. Introducing   two additional scalar fields besides $\la$ in DFTs, one can see that after DFTs the pressureless mimetic matter turns out to be moving arbitrarily (i.e.,   without the potentiality restriction) \cite{statja48}.
			Another interesting way to modify DFT   in Equation      \eqref{v1} is the consideration of the so-called disformal transformations \cite{arXiv1407.0825}, which were extensively studied in the context of scalar-tensor theories as well as other aspects of gravity.

			\subsection{Disformal Transformations}\label{4.2}
			For the  first time, the disformal transformation
			in gravity was considered by Bekenstein \cite{gr-qc/9211017}. These can be written in the following form:
			\begin{align}\label{disf}
			g_{\mu\nu} = A(\la,X) \tilde{g}_{\mu\nu}+B(\la,X) \dd_\mu \la\, \dd_\nu \la,
			\end{align}
			where $X\equiv \tilde{g}^{\mu\nu} \dd_\mu \la\, \dd_\nu \la$.
			Note that in the general case there is an increase of the number of independent variables:
			ten components of the physical metrics $g_{\m\n}$ transforms into  eleven new degrees of freedom, namely ten components of auxiliary metric $\tilde g_{\m\n}$ and one scalar field $\la$.
			The DFT corresponding to mimetic gravity  (Equation      \eqref{v1}) is a singular case of Equation      \eqref{disf}, in which
			\begin{align}\label{q0}
			A(\la,X)=X,\qquad B(\la,X)=0,
			\end{align}
			thus the number of new independent variables in fact lowers by one due to the presence of \mbox{Weyl symmetry.}
			
			The extension of dynamics after an arbitrary disformal transformation depends on the question whether it is possible to obtain an arbitrary variation of the physical metric
			\begin{align}\label{q1}
			\de g_{\m\n}=\Psi_{\m\n}{}^{\al\be}(\la,X,\tilde g_{\m\n})\de \tilde g_{\al\be}
			\end{align}
			by choosing the variation of the auxiliary metric $\de \tilde g_{\m\n}$
			{(see the discussion in    Section \ref{2.1})}. It  in turn  depends on the invertibility of the quantity $\Psi_{\m\n}{}^{\al\be}(\la,X,\tilde g_{\m\n})$ (its explicit form is quite  bulky, see, e.g.,~\cite{arXiv1407.0825}).
			The analysis of the EoM of the theory, appearing as a result of the DFT   in Equation      \eqref{disf} in action in   Equation      \eqref{EHM}
			shows \cite{arXiv1407.0825} that the extension of dynamics does not occur (i.e.,   the EoM are equivalent to the Einstein equations)
			if the functions $A$ and $B$ are chosen in a way that for all $\la$ and $X$
			\begin{align}\label{det}
			\frac{\partial (B+A/X)}{\partial X} A \neq 0,
			\end{align}
			i.e., for generic disformal transformations.
			As can be seen, for the mimetic transformation, when the functions $A$ and $B$ are defined by Equation      \eqref{q0}, the condition  in  Equation      \eqref{det} does not hold,
			which corresponds to the extension of dynamics in this case.
			
			
			\section{Concluding Remarks}\label{Disc}
			In the above sections, we   show examples of extension, restriction and conservation of dynamics in various theories
			after a differential transformation of the fields in the action.
			To determine the behavior of a theory after DFT is applied to the action, one should begin with two following questions:
			\begin{enumerate}
				\item Is it possible to attain an arbitrary value of the original variables by choosing new variables?
				\item Is it possible to attain an arbitrary value of the {\em variations} of original variables by choosing the {\em variations} of new variables?
			\end{enumerate}
			
			Note that the fixation of the endpoints is assumed for the variations of variables. We also assume that the original variables do not possess any unphysical degrees of freedom, i.e., gauge ones.

			If the answer to   both questions is positive, then the dynamics is conserved. Examples of this situation are the second mechanical model from   Section \ref{2.1} (see Equations      \eqref{meh2-1} and       \eqref{meh2-2}) and the general case of disformal transformation for
			gravity from    Section \ref{4.2}.

			If the answer to the first question is negative, then the dynamics might become restricted. However, in some cases,  this does not occur. For instance, the restriction of dynamics does occur in generalized Hilbert--Palatini approach mentioned at the end of    Section \ref{2.3}, when the independent connection is not assumed to be symmetric \cite{statja47}. However, when we perform the same DFT in the usual Hilbert--Palatini approach (see Section  \ref{2.3}), the restriction of dynamics does not occur,  thus we have an \mbox{equivalent formulation}.
			
			If the second question is answered negatively,
			then an extension of dynamics is possible. \mbox{It can} be seen clearly in the toy example from    Section \ref{2.1}, and the same situation takes place in \mbox{Sections \ref{2.2}, \ref{secIII} and \ref{4.1}}.
			However, in some cases, an extension of dynamics can also be absent. The DFT in the usual Hilbert--Palatini approach from    Section \ref{2.3} can serve as the example of such situation. It can be said that in this example negative answers to the first and second questions somewhat compensate each other.
		}
		
		The simplest part in the analysis of a given DFT is to determine how the number of fields change after it.
		If the number of fields has decreased (i.e.,   the number of new fields are smaller than the number of the original ones), then the answer to   both questions is obviously negative.
		It takes place for the DFTs  considered in    Section \ref{2.3}.
		If the number of fields has increased, it can be interpreted as the appearance of the gauge invariance in the theory, since the change of some new independent variables does not affect the action. It takes place for the  DFTs considered in    Section \ref{secIV}.
		The especially simple situation takes place for the DFT in Equation      \eqref{v1}, when the number of new variables is formally 11, but one of them is a conformal mode of
		the auxiliary metric $\tilde g_{\m\n}$ and turns out to be unphysical DoF (e.g.,   gauge one).
		In the remaining cases, the number of fields is conserved.
		It should be noted that, \mbox{in the} cases when the number of fields is {\em not} decreasing, the answers to the above questions are not a priori known,  thus they should be addressed case by case.

		In the present paper, we restrict  ourselves to the consideration of the effects of DFTs in the action,
		since in this way it is easier to achieve the extension of dynamics. A more detailed consideration of the effects and issues appearing after the substitution of fields within the action can be found in \cite{0909.4151}, although the extension of dynamics is not the main topic of this paper. Some theorems related to the field transformation within the action can also be found in \cite{1702.01849}.
		
		On the contrary, when one transforms the equations of motion into another form  that can be easier to solve, one can be interested in conservation of dynamics.
		{In the particular case when the number of variables is conserved},
		one can pose the question of invertibility of {DFTs}.
		This question is studied in detail  in   \cite{1907.12333}, where the authors employed the method of characteristics to obtain the necessary and sufficient conditions for the invertibility of a given DFT.
		
		As can be seen, differential transformations of  fields often give  the theory some interesting properties, which can be of use in the theory of gravity and its cosmological applications and  has drawn significant attention in the recent years. In this paper we put together the main results in this field. However, a detailed review of the general behavior of a theory after DFTs is a laborious task and lays beyond the scope of our present paper.
		
		\vspace{6pt}
		
		\bf{Acknowledgments}}. {The authors are grateful to Roman Ilin and to Alexey Golovnev for the useful references. The work of S. P. and A. S. was supported by RFBR Grant No. 20-01-00081.}


	\end{document}